\documentclass[conference]{IEEEtran}

\ifCLASSINFOpdf
\else
\fi

\usepackage{amsmath, amssymb, bm, cite, epsfig}
\usepackage{epstopdf}
\usepackage{graphicx}
\usepackage{array}
\usepackage{multirow}
\renewcommand{\arraystretch}{1.5}
\usepackage{amsfonts}
\usepackage{tabularx} 
\usepackage{tabu}
\usepackage{pbox}
\usepackage{makecell}

\usepackage{booktabs}
\usepackage{ctable}
\usepackage{xcolor,colortbl}
\definecolor{Gray}{gray}{0.85}
\definecolor{LightCyan}{rgb}{0.8,1,1}
\def\beq{\begin{equation}}
\def\eeq{\end{equation}}
\def\beqa{\begin{eqnarray}}
\def\eeqa{\end{eqnarray}}
\def\beqan{\begin{eqnarray*}}
\def\eeqan{\end{eqnarray*}}

\def\PL{\mathrm{PL}}
\def\dB{\mathrm{dB}}

\def\tm1{t\! - \! 1}
\def\tp1{t\! + \! 1}

\def\PL{\mathrm{PL}}
\def\dB{\mathrm{dB}}
\def\PLE{\mathrm{PLE}}
\def\FSPL{\mathrm{FSPL}}
\def\log{\mathrm{log}}
\def\ABG{\mathrm{ABG}}
\def\CI{\mathrm{CI}}

\usepackage{tikz}
\usetikzlibrary{calc}

\begin{document}
\title{Propagation Path Loss Models for 5G Urban  Micro- and Macro-Cellular Scenarios}


\author{\IEEEauthorblockN{Shu Sun$^{a*}$, Theodore S. Rappaport$^a$, Sundeep Rangan$^a$, Timothy A. Thomas$^b$, Amitava Ghosh$^b$, Istv$\acute{a}$n Z. Kov$\acute{a}$cs$^{c}$,} {Ignacio Rodriguez$^{d}$, Ozge Koymen$^e$, Andrzej Partyka$^e$, and Jan J{\"a}rvel{\"a}inen$^f$}
\IEEEauthorblockA{$^a$NYU WIRELESS and Tandon School of Engineering, New York University, Brooklyn, NY, USA 11201\\
$^b$Nokia, Arlington Heights, IL, USA 60004\\
$^{c}$Nokia, Aalborg, Denmark 9220\\
$^{d}$Aalborg University, Aalborg, Denmark 9220\\
$^{e}$Qualcomm R\&D, Bridgewater, NJ, USA, 08807\\
$^{f}$Aalto University School of Electrical Engineering, Espoo, Finland, FI-00076\\
$^*$Corresponding author: ss7152@nyu.edu }

\thanks{Sponsorship for this work was provided by the NYU WIRELESS Industrial Affiliates program and NSF research grants 1320472, 1302336, and 1555332.}
}

\maketitle
\begin{tikzpicture}[remember picture, overlay]
\node at ($(current page.north) + (0,-0.25in)$) {S. Sun~\textit{et al.}, \rq\rq{}Propagation Path Loss Models for 5G Urban  Micro- and Macro-Cellular Scenarios,\rq\rq{}};
\node at ($(current page.north) + (0,-0.4in)$) {in \textit{2016 IEEE 83rd Vehicular Technology Conference (VTC2016-Spring)}, May 2016.};
\end{tikzpicture}

\begin{abstract}
This paper presents and compares two candidate large-scale propagation path loss models, the alpha-beta-gamma (ABG) model and the close-in (CI) free space reference distance model, for the design of fifth generation (5G) wireless communication systems in urban micro- and macro-cellular scenarios. Comparisons are made using the data obtained from 20 propagation measurement campaigns or ray-tracing studies from 2 GHz to 73.5 GHz over distances ranging from 5 m to 1429 m. The results show that the one-parameter CI model has a very similar goodness of fit (i.e., the shadow fading standard deviation) in both line-of-sight and non-line-of-sight environments, while offering substantial simplicity and more stable behavior across frequencies and distances, as compared to the three-parameter ABG model. Additionally, the CI model needs only one very subtle and simple modification to the existing  3GPP floating-intercept path loss model (replacing a constant with a close-in free space reference value) in order to provide greater simulation accuracy, more simplicity, better repeatability across experiments, and higher stability across a vast range of frequencies.
\end{abstract}

\IEEEpeerreviewmaketitle
\section{Introduction}
The rapid growth of personal communication devices such as smart phones and tablets, and consumer demand for ubiquitous data access, have motivated carriers to provide higher data rates and quality. Innovative technologies and new frequency bands such as millimeter waves (mmWaves) are needed to meet this impending demand\cite{Rap13:Access}, driving the development of the fifth generation (5G) wireless communications\cite{Han16:VTC}. Emerging 5G communication systems are expected to introduce revolutionary technologies, while utilizing potential new spectra and novel architectural concepts\cite{And14,Boc14}, hence it is critical to develop new standards and channel models to assist engineers in system design. Channel characterization at mmWave frequencies has been conducted by prior researchers. Violette \textit{et al}. studied wideband non-line-of-sight (NLOS) channels at 9.6, 28.8, and 57.6 GHz in downtown Denver\cite{Vio88}; Outdoor propagation measurements and modeling at the 60 GHz band were carried out in various city streets\cite{Lov94, Smu97}; Samsung has been active in measuring and modeling mmWave channels for future mobile communications\cite{Hur14,Roh14}; Channel measurements at 81 GHz to 86 GHz of the E-band were performed by Aalto University for point-to-point communications in a street canyon scenario in Helsinki, Finland\cite{Kyro12}; Extensive propagation measurements have been performed at 28 GHz, 38 GHz, and 73 GHz in urban micro-cellular (UMi), urban macro-cellular (UMa), and/or indoor scenarios\cite{Rap13:Access,Rap15:TCOM,Rap13:TAP}, from which spatial and temporal statistics were extracted in combination with the ray-tracing technique. Omnidirectional path loss models in dense urban environments at 28 GHz and 73 GHz were investigated in\cite{Mac14:PIMRC}. There are numerous other measurement campaigns throughout the world at mmWave frequencies that are being or have just been performed and have not yet been published, such as the measurement data provided in this paper. 


This paper presents the alpha-beta-gamma (ABG) and close-in (CI) free space reference distance path loss models\cite{GRM13:Globecom,Pie12,And95} at mmWave frequencies, and provides a head-to-head comparison between the parameters and shadow fading (SF) standard deviations in these two models in both UMi and UMa scenarios, using 20 sets of measurement or ray-tracing data contributed by New York University (NYU), Nokia, Aalborg University (AAU), Qualcomm, and Aalto University. 

\section{Close-In Reference Distance and Alpha-Beta-Gamma Path Loss Models}
Both ABG and CI path loss models are generic all-frequency models that describe large-scale propagation path loss at all relevant frequencies in a certain scenario. The CI model is easily implemented in existing 3GPP models by making a very subtle modification --- by replacing a floating non-physically based constant with a frequency-dependent constant that represents free space path loss in the first meter of propagation.
The equation for the ABG model is given by~\eqref{ABG1}:
\begin{equation}\label{ABG1}
\begin{split}
\PL^{\ABG}(f,d)[\dB]=&10\alpha \log_{10}\left(\frac{d}{1~m}\right)+\beta\\
&+10\gamma \log_{10}\left(\frac{f}{1~GHz}\right)+\chi_{\sigma}^{\ABG}
\end{split}
\end{equation}

\noindent where $\PL^{\ABG}(f,d)$ denotes the path loss in dB over frequency and distance, $\alpha$ and $\gamma$ are coefficients showing the dependence of path loss on distance and frequency, respectively, $\beta$ is an optimized offset value for path loss in dB, $d$ is the 3D transmitter-receiver (T-R) separation distance in meters, $f$ is the carrier frequency in GHz, and $\chi_{\sigma}^{\ABG}$ is the SF standard deviation describing large-scale signal fluctuations about the mean path loss over distance. Note that the ABG model, when used at a single frequency, reverts to the floating-intercept model with two parameters with $\gamma$ set to 0 or 2\cite{WINNER,Rap15:TCOM,GRM13:Globecom}. The coefficients $\alpha$, $\beta$, and $\gamma$ are obtained from measured data using the closed-form solutions that minimize the SF standard deviation given in the Appendix. 

The equation for the CI model is given in~\eqref{CI1}:
\begin{equation}\label{CI1}
\PL^{\CI}(f,d)[\dB]=\FSPL(f, 1~m)[\dB]+10n\log_{10}\left(d\right)+\chi_{\sigma}^{\CI}
\end{equation}

\noindent where $n$ denotes the single model parameter, the path loss exponent (PLE), with 10$n$ describing path loss in dB in terms of decades of distances beginning at 1 m (making it very easy to compute power over distance without a calculator), $d$ is the 3D T-R separation distance, and $\FSPL(f,1~m)$ denotes the free space path loss in dB at a T-R separation distance of 1 m at the carrier frequency $f$:
\begin{equation}\label{FSPL1}
\FSPL(f,1~m)[\dB]=20\log_{10}\left(\frac{4\pi f}{c}\right)
\end{equation}

\noindent where $c$ is the speed of light. Note that the CI model inherently has an intrinsic frequency dependency of path loss embedded within the 1 m free space path loss value, and it has only one parameter, PLE, to be optimized, as opposed to three parameters in the ABG model ($\alpha, \beta$, and $\gamma$). The optimized minimum error CI PLE parameter (see the Appendix) is found by first subtracting the 1 m FSPL value from each path loss data point to obtain $n$\cite{Rap15:TCOM,Rappaport:Wireless2nd,Tho16:VTC}. The CI model can then be applied across a vast range of frequencies using~\eqref{CI1} and the single value of $n$ that is very stable across a wide range of frequencies. The ABG model also is applied across a vast range of frequencies using its three parameters, but the floating point parameters vary substantially across different frequencies\cite{Rap15:TCOM,Tho16:VTC}, meaning the ABG model will have more error when extrapolating the model outside of the frequencies or distances that data was used to determine parameters. 

Both the ABG~\eqref{ABG1} and CI~\eqref{CI1} path loss models are a function of both distance and frequency, where the CI model has its frequency dependence expressed primarily by the frequency-dependent FSPL term~\eqref{FSPL1} in the first meter of propagation. While the ABG model offers some physical basis in the $\alpha$ term, being based on a 1 m reference distance similar to the $n$ term in~\eqref{CI1}, it departs from physics when introducing both an offset $\beta$ (which is basically an optimization parameter that is not physically based), and a frequency weighting term $\gamma$ which has no proven physical basis, although recent indoor measurements show that the path loss increases with frequency across the mmWave band\cite{Deng15} (both of these parameters are basically used for curve fitting, as was done in the WINNER floating-intercept (\textit{alpha-beta}, or AB) model)\cite{WINNER,Rap15:TCOM,GRM13:Globecom}. It is noteworthy that the ABG model is identical to the CI model if we equate $\alpha$ in the ABG model in~\eqref{ABG1} with the PLE $n$ in the CI model in~\eqref{CI1}, $\gamma$ in~\eqref{ABG1} with the free space PLE of 2, and $\beta$ in~\eqref{ABG1} with $20\log_{10}(4\pi/c)$  in~\eqref{FSPL1}. 

\begin{table}
\renewcommand{\arraystretch}{1.0}
\begin{center}
\caption{Parameters in the ABG and CI path loss models in UMi and UMa scenarios. SC denotes street canyon, OS means open square, Freq. Range represents frequency range, and Dist. Range denotes distance range.}~\label{tbl:UMi_UMa}
\begin{tabular}{|c|c|c|c|c|c|c|c|c|}
\hline 
 Sce. & Env.& \makecell{Freq.\\Range\\(GHz)}&\makecell{Dist. \\Range\\(m)}&Model & \makecell{$\PLE$\\/$\alpha$} & \makecell{$\beta$ \\($\dB$)} & $\gamma$ & \makecell{$\sigma$ \\($\dB$)} \\ \cline{1-9}
\multirow{4}{*}{\makecell{UMi\\SC}} & \multirow{2}{*}{LOS}& \multirow{2}{*}{2-73.5}& \multirow{2}{*}{5-121} & ABG & 2.0 & 31.4 & 2.1 & 2.9 \\ \cline{5-9}
 & & & & CI & 2.0 & - & - & 2.9 \\ \cline{2-9}
 & \multirow{2}{*}{NLOS}  & \multirow{2}{*}{2-73.5}  & \multirow{2}{*}{19-272} & ABG & 3.5 & 24.4 & 1.9 & 8.0 \\ \cline{5-9}
  & & & & CI & 3.1 & - & - & 8.1 \\ \cline{1-9}
\multirow{4}{*}{\makecell{UMi\\OS}} & \multirow{2}{*}{LOS}& \multirow{2}{*}{2-60}& \multirow{2}{*}{5-88} & ABG & 2.6 & 24.0 & 1.6 & 4.0 \\ \cline{5-9}
  & & & & CI & 1.9 & - & - & 4.7 \\ \cline{2-9}
 & \multirow{2}{*}{NLOS}& \multirow{2}{*}{2-60}& \multirow{2}{*}{8-235} & ABG & 4.4 & 2.4 & 1.9 & 7.8 \\ \cline{5-9}
  & & & & CI & 2.8 & - & - & 8.3 \\ \cline{1-9}
\multirow{4}{*}{UMa} & \multirow{2}{*}{LOS} & \multirow{2}{*}{2-73.5} & \multirow{2}{*}{58-930} & ABG & 2.8 & 11.4 & 2.3 & 4.1 \\ \cline{5-9}
 & &  & & CI & 2.0 & - & - & 4.6 \\ \cline{2-9}
 & \multirow{2}{*}{NLOS}& \multirow{2}{*}{2-73.5}& \multirow{2}{*}{45-1429} & ABG & 3.3 & 17.6 & 2.0 & 9.9 \\ \cline{5-9}
 & &  & & CI & 2.7 & - & - & 10.0 \\ \cline{1-9}
\end{tabular}
\end{center}
\end{table}

\begin{figure}
\centering
 \includegraphics[width=2.9in]{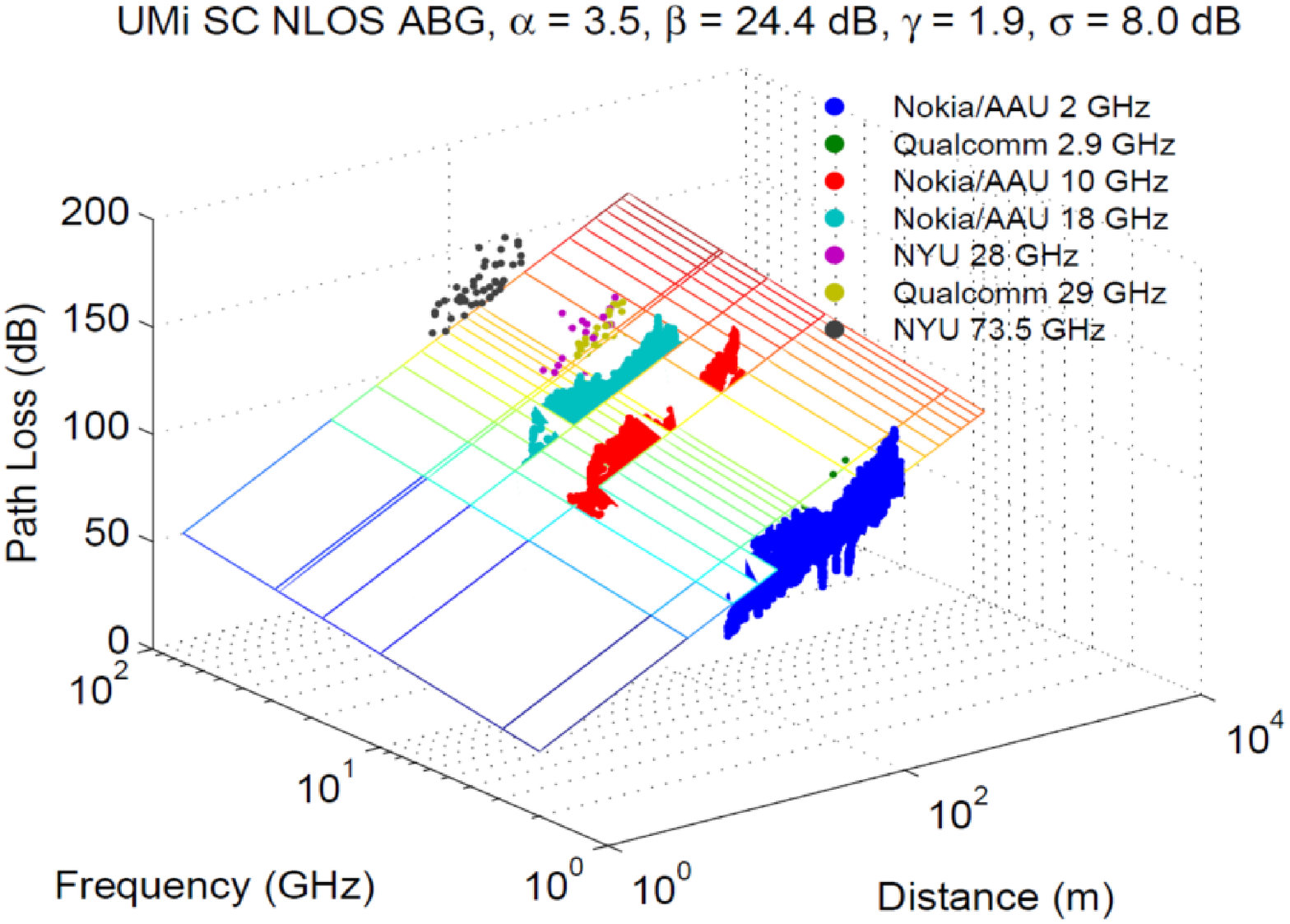}
    \caption{ABG path loss model in the UMi SC scenario across different frequencies and distances in NLOS environments.}
    \label{fig:UMi_SC_NLOS_ABG}
\end{figure}

\begin{figure}
\centering
 \includegraphics[width=2.9in]{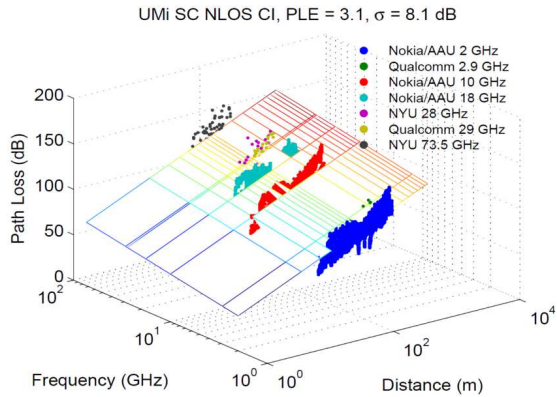}
    \caption{CI path loss model in the UMi SC scenario across different frequencies and distances in NLOS environments.}
    \label{fig:UMi_SC_NLOS_CI}
\end{figure}

The CI model is based on fundamental principles of wireless propagation, dating back to Friis and Bullington, where the PLE offers insight into path loss based on the environment, having a value of 2 in free space as shown by Friis and a value of 4 for the asymptotic two-ray ground bounce propagation model\cite{Rappaport:Wireless2nd}. Previous
UHF (Ultra-High Frequency)/microwave models used a close-in reference distance of 1 km or 100 m since base station towers were tall without any nearby obstructions and inter-site distances were on the order of many kilometers for those frequency bands\cite{Rappaport:Wireless2nd,Hata:TVT80}. We use $d_0$ = 1 m in mmWave path loss models since base stations will be shorter or mounted indoors, and closer to obstructions~\cite{Rap13:Access,Rap15:TCOM}. The CI 1 m reference distance is a conveniently suggested standard that ties the true transmitted power or path loss to a convenient close-in distance of 1 m, as suggested in\cite{Rap15:TCOM}. Standardizing to a reference distance of 1 m makes comparisons of measurements and models simpler, and provides a standard definition for the PLE, while enabling intuition and rapid computation of path loss without a calculator. Emerging mmWave mobile systems will have very few users within a few meters of the base station antenna, and close-in users in the near field will have strong signals or will be power-controlled compared to typical users much farther from the transmitter such that any path loss error in the near-field (between 1 m and the Fraunhofer distance) will be so much smaller than the dynamic range of signals experienced by users in a commercial system. Additionally, the 1 m CI model also offers more accurate prediction on path loss beyond measurement ranges when compared to the AB and ABG models as shown in\cite{Tho16:VTC,Akd14,Rap15:TCOM}. 

\begin{table*}
\renewcommand{\arraystretch}{1.0}
\begin{center}
\caption{Parameters in the ABG and CI path loss models in UMi street canyon (SC) scenario in NLOS environments (Env.) for different frequency (Freq.) and distance (Dist.) ranges. M denotes measurement data, while R means ray-tracing data.}~\label{tbl:UMi_SC}
\begin{tabu}{|c|c|c|c|c|c|c||c|[1.7pt black]c|c|c|[1.7pt black]c|[1.7pt black]c|[1.7pt black]c|}
\hline 
 Sce. & Env.&\makecell{Freq./Freq. \\Range (GHz)} & Company &\makecell{\# of\\Data Points}&\makecell{Dist. Range\\(m)} & Type& $n^{\CI}$& $\alpha^{\ABG}$ & \makecell{$\beta^{\ABG}$ \\($\dB$)} & $\gamma^{\ABG}$ &\makecell{$\sigma^{\CI}$ \\($\dB$)} & \makecell{$\sigma^{\ABG}$ \\($\dB$)} & \makecell{$\sigma^{\CI}-\sigma^{\ABG}$ \\($\dB$)}\\ \cline{1-14}
\multirow{9}{*}{\makecell{UMi\\ SC}} & \multirow{9}{*}{NLOS} & 2 & Nokia/AAU &27158& 19-272 & M &  3.1 &3.5 & 25.0 & 2 & 7.7 & 7.6 & 0.1\\ \cline{3-14}
 & & 2.9 & Qualcomm &34&109-235 & M & 2.9 & 3.9 & 10.2 & 2 & 3.3 &3.2 & 0.1\\ \cline{3-14}
 & & 18 & Nokia/AAU&13934&19-272 & M & 3.1 & 3.5 & 24.0 & 2 & 8.0 &8.0 & 0.0\\ \cline{3-14}
 & & 28 & NYU &20& 61-186 & M & 3.4 & 2.5 & 51.7 & 2 & 9.7 & 9.7 & 0.0\\ \cline{3-14}
& & 29 & Qualcomm&34&109-235 & M & 3.2 &4.2 & 11.0 & 2 & 5.4 &5.3 & 0.1\\ \cline{3-14}
& & 73.5 & NYU&53& 48-190 & M & 3.4 &2.9 & 43.2 & 2 & 7.9 &7.8 & 0.1 \\  \tabucline[1.7pt black off 0pt]{3-14}
 & &2-18 &- &54350&19-272 & M & 3.1 &3.5 &24.4 & 1.9 & 8.1 &8.0 &0.1\\ \cline{3-14}
& & 28-73.5 & - &107&48-235 & M & 3.3 & 2.7 & 36.1 & 2.6 & 8.0 & 7.8 & 0.2 \\ \cline{3-14}
& & 2-73.5 & - & 54457&19-272 & M & 3.1 & 3.5 & 24.4 & 1.9 & 8.1 & 8.0 & 0.1\\ \cline{1-14}
\end{tabu}
\end{center}
\end{table*} 

\begin{table*}
\renewcommand{\arraystretch}{1.0}
\begin{center}
\caption{Parameters in the ABG and CI path loss models in UMi open square (OS) scenario in NLOS environments (Env.) for different frequency (Freq.) ranges and distance (Dist.) ranges. M denotes measurement data, while R means ray-tracing data.}~\label{tbl:UMi_OS}
\begin{tabu}{|c|c|c|c|c|c|c||c|[1.7pt black]c|c|c|[1.7pt black]c|[1.7pt black]c|[1.7pt black]c|}
\hline 
 Sce. & Env.&\makecell{Freq./Freq. \\Range (GHz)}& Company&\makecell{\# of\\Data Points}& \makecell{Dist. Range\\(m)} & Type& $n^{\CI}$& $\alpha^{\ABG}$ & \makecell{$\beta^{\ABG}$ \\($\dB$)} & $\gamma^{\ABG}$ &\makecell{$\sigma^{\CI}$ \\($\dB$)} & \makecell{$\sigma^{\ABG}$ \\($\dB$)} & \makecell{$\sigma^{\CI}-\sigma^{\ABG}$ \\($\dB$)}\\ \cline{1-14}
\multirow{8}{*}{\makecell{UMi\\ OS}} & \multirow{8}{*}{NLOS} & 2 & Nokia/AAU &10377&17-138 & M &  2.9 & 4.7 & -2.2 & 2 & 7.9 & 7.4 & 0.5\\ \cline{3-14}
 & & 2.9 & Qualcomm &34& 109-235 & M & 2.9 & 3.9 & 10.2 & 2 & 3.3 &3.2 & 0.1\\ \cline{3-14}
 & & 18 & Nokia/AAU &6073&23-138 & M & 2.8 & 4.9 & -7.7 & 2 & 8.7 &7.9 & 0.8\\ \cline{3-14}
& & 29 & Qualcomm &34& 109-235 & M & 3.2 &4.2 & 11.0 & 2 & 5.4 &5.3 & 0.1 \\ \cline{3-14}
& & 60 & Aalto &246& 8-36 & M & 3.2 &2.2 & 46.5 & 2 & 2.2 &1.8 & 0.4\\ \tabucline[1.7pt black off 0pt]{3-14}
 & & 2-18 & - &21888&17-235 & M & 2.8 &4.7 & -3.1 & 1.8 & 8.3 &7.6 & 0.7\\ \cline{3-14}
& & 29-60 & - &280& 8-235 & M & 3.2 &2.4 & 74.2 & 0.3 & 2.8 &2.6 & 0.2 \\ \cline{3-14}
& & 2-60 & - &22168&8-235 & M & 2.8 &4.4 & 2.4 & 1.9 & 8.3 &7.8 & 0.5\\ \cline{1-14}
\end{tabu}
\end{center}
\end{table*} 

\begin{table*}
\renewcommand{\arraystretch}{1.0}
\begin{center}
\caption{Parameters in the ABG and CI path loss models in UMa scenario in NLOS environments (Env.) for different frequency (Freq.) ranges and distance (Dist.) ranges. M denotes measurement data, while R means ray-tracing data.}~\label{tbl:UMa}
\begin{tabu}{|c|c|c|c|c|c|c||c|[1.7pt black]c|c|c|[1.7pt black]c|[1.7pt black]c|[1.7pt black]c|}
\hline 
 Sce. & Env.& \makecell{Freq./Freq. \\Range (GHz)} & Company&\makecell{\# of\\Data Points}& \makecell{Dist. Range\\(m)} & Type& $n^{\CI}$& $\alpha^{\ABG}$ & \makecell{$\beta^{\ABG}$ \\($\dB$)} & $\gamma^{\ABG}$ &\makecell{$\sigma^{\CI}$ \\($\dB$)} & \makecell{$\sigma^{\ABG}$ \\($\dB$)} & \makecell{$\sigma^{\CI}-\sigma^{\ABG}$ \\($\dB$)}\\ \cline{1-14}
\multirow{10}{*}{UMa} & \multirow{10}{*}{NLOS} & 2 & Nokia/AAU &69542& 45-1429 & M, R & 2.7 &3.6 & 7.6 & 2 & 9.6 & 9.4 & 0.2\\ \cline{3-14}
 &  & 10.25 & Nokia &16743& 45-1174 & R & 2.7 & 2.2 & 47.6 & 2 & 12.6 &12.5 & 0.1\\ \cline{3-14}
 & & 18 & Nokia/AAU &27154& 90-1429 & M & 2.9 & 3.7 & 8.0 & 2 & 6.5 & 6.1 & 0.4\\ \cline{3-14}
& & 28.5 & Nokia&16416& 45-1174 & R & 2.7 &1.9 & 52.3 & 2 & 12.1 &12.0 & 0.1\\ \cline{3-14}
& & 37.625 & NYU &12&61-377 & M & 2.7 &1.0 & 69.4 & 2 & 10.5 &9.6 & 0.9\\ \cline{3-14}
& & 39.3 & Nokia &16244&45-1174 & R & 2.6 & 1.8 & 53.8 & 2 & 11.7 & 11.6 & 0.1\\ \cline{3-14}
& & 73.5 & Nokia &15845&45-1174 & R & 2.6 &1.9 & 49.7 & 2 & 10.1 &10.0 & 0.1\\ \tabucline[1.7pt black off 0pt]{3-14}
 & & 2-18 & - &137981&45-1429 & M, R & 2.7 &3.6 & 7.4 & 2.4 & 9.2 &8.9 & 0.3\\ \cline{3-14}
& & 28.5-73.5 & - &48517&45-1174 & M, R & 2.6 & 1.9 & 64.6 & 1.2 & 11.4 & 11.2 & 0.2 \\ \cline{3-14}
& & 2-73.5 & - &186498&45-1429 & M, R & 2.7 & 3.3 & 17.6 & 2.0 & 10.0 & 9.9 & 0.1\\ \cline{1-14}
\end{tabu}
\end{center}
\end{table*} 

Using the two path loss models described above, and the measurement and ray-tracing data over wide ranges of mmWave frequencies (2 to 73.5 GHz) and distances (5 to 1429 m) from the companies and universities across the world as mentioned above, we computed the path loss parameters for the two models. Both the PLE in the CI model and the $\alpha$, $\beta$, and $\gamma$ parameters in the ABG model were calculated via the MMSE fit on all of the combined path loss data from all measured frequencies and distances, using closed-form solutions that minimize the SF standard deviation, as detailed in the Appendix. All of the scattered path loss data samples were used in the analysis without additional local averaging, which is reasonable as long as both models were compared using identical data processing methods (note that the data from some campaigns had been averaged over small local distances before being provided for use in this paper). In addition, all path loss values were upper-bounded to 180 dB based on reasonable assumptions for typical high gain steerable antennas and 1 W transmit power levels, as well as the realistic sensitivity of the receivers in real-world measurement systems\cite{Rap15:TCOM}. 

\begin{figure}
\centering
 \includegraphics[width=2.6in]{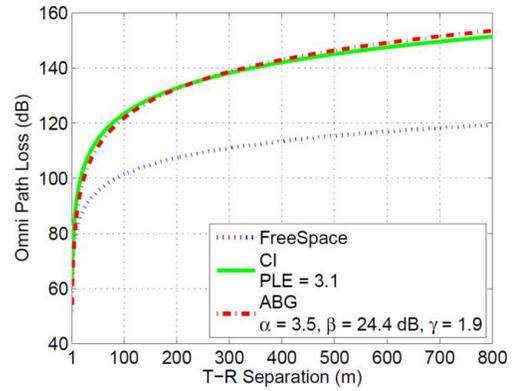}
    \caption{Example comparison of free space, CI and ABG path loss models at 28 GHz for UMi street canyon NLOS environments using the parameters derived with measurements from 2 - 73.5 GHz in Table~\ref{tbl:UMi_SC}.}
    \label{fig:FS_CI_ABG}
\end{figure}

Figs.~\ref{fig:UMi_SC_NLOS_ABG} and~\ref{fig:UMi_SC_NLOS_CI} show scatter plots of all the data sets optimized for the ABG and CI models in the UMi street canyon (SC) scenario in NLOS environments, respectively. 
Table~\ref{tbl:UMi_UMa} summarizes the path loss parameters in the ABG and CI models for both the UMi and UMa scenarios in both LOS and NLOS environments. As shown by Table~\ref{tbl:UMi_UMa}, the CI model provides PLEs of 2.0 and 1.9 in LOS environments, which agree well with a free space PLE of 2. Although the CI model yields slightly higher (by up to 0.7 dB) SF standard deviation than the ABG model, it is within standard measurement error arising from frequency and temperature drift, connector and cable flex variations, and calibration errors in an actual measurement campaign, and are within the practical error of ray tracing anomalies such as imperfect databases or double ray counting. 
Tables~\ref{tbl:UMi_SC},~\ref{tbl:UMi_OS}, and~\ref{tbl:UMa} list the modeling parameters in the ABG and CI models at different frequencies for the UMi SC scenario, UMi OS scenario, and UMa scenario in NLOS environments, respectively. Note that  for single frequencies, $\gamma$ in the ABG model is set to 2, thus reverting to the AB model used in 3GPP and WINNER II channel models\cite{WINNER,3GPP:25996,3GPP:36814}. The parameter values in the last row and the 3rd row from last in Table~\ref{tbl:UMi_SC} are very similar, the reason may be that the data points for 2-18 GHz account for a major proportion among the data points for the entire 2-73.5 GHz, as shown by the 5th column in Table~\ref{tbl:UMi_SC}. 
Fig.~\ref{fig:FS_CI_ABG} illustrates an intuitive comparison of the CI and ABG models with the free space path loss line at 28 GHz for UMi street canyon NLOS environments, using the parameters derived from all data from 2 - 73.5 GHz in Table~\ref{tbl:UMi_SC}. Note that the ABG model in this NLOS scenario gives physically unrealistic path loss values (much less than the free space) at very close distances (out to 4 m), and underestimates signal strength (compared to CI) at very large distances. 
The main observations from these figures and tables are as follows:
\begin{itemize}
\item The $\alpha$ and $\beta$ parameters in the AB model can vary widely, as much as 2.7 and 61.8 dB across frequencies, respectively, as shown in Table~\ref{tbl:UMa}. The wild variation of $\alpha$ and $\beta$ in the AB model was also observed in\cite{Rap15:TCOM}. The parameters in the ABG model also vary wildly over all frequencies and distances. For example, as illustrated by the last three rows in Tables~\ref{tbl:UMi_SC},~\ref{tbl:UMi_OS}, and~\ref{tbl:UMa}, the largest variation in $\alpha$, $\beta$, and $\gamma$ are 0.8, 11.7 dB, and 0.7 for UMi SC, respectively, 2.3, 77.3 dB, and 1.6 for UMi OS, respectively, 1.7, 57.2 dB, and 1.2 for UMa, respectively. These are huge variations, and show how sensitive and prone to error the ABG model may be without having a continuum of data from all frequencies, distances, and possible TX/RX locations.
\item The PLE $n$ in the CI model varies only marginally for both single frequency and multiple frequency cases, with a largest variation of merely 0.2, 0.4, and 0.1 for UMi SC, UMi OS, and UMa, respectively, in the multiple frequency case, as shown by the last three rows in Tables~\ref{tbl:UMi_SC},~\ref{tbl:UMi_OS}, and~\ref{tbl:UMa}. Single frequency UMi data in Tables~\ref{tbl:UMi_SC} and~\ref{tbl:UMi_OS} show the PLE tends to increase only slightly with an increase in frequency, as suggested in\cite{Rap15:TCOM}, and Table~\ref{tbl:UMa} shows no significant frequency sensitivity for the PLE with taller UMa transmitters.
\item The SF standard deviations for the CI and ABG models differ by only a fraction of a dB in most cases over all frequencies and distances, always less than an order of magnitude of the SF standard deviation and typically within 0.2 dB, with a largest difference of only 0.9 dB (where the standard deviation for both models in that case is more than 9 dB). It is important to note that the difference in SF between the CI and ABG models is always less than an order of magnitude of the SF for either model, making the models virtually identical in accuracy over frequency and distance.
\item As shown in Fig.~\ref{fig:FS_CI_ABG}, the parameters derived from 2 to 73.5 GHz for UMi street canyon NLOS environments, when applied at 28 GHz, indicate that the ABG model underestimates path loss to be less than free space when very close to the TX, and the CI model overestimates path loss close to the transmitter when compared to the ABG model, yet this is where errors are not as important in practical system design\cite{Rap15:TCOM}. More importantly, the ABG model overestimates path loss (i.e., underestimates interference at greater distances) compared with the CI model. Thus, the ABG model could underestimate the true interference in system design, while the CI model is more safe and conservative when used to analyze interference-limited systems.
\end{itemize}


\section{Conclusion}
In this paper, we provided a comparison of the ABG and CI path loss models in the mmWave frequency bands, using measured data and ray-tracing from 2 GHz to 73.5 GHz obtained from 20 data sets from research groups across the world. The CI model is physically tied to the transmitter power using a close-in free space reference, and standardizes all measurements around an inherent 1 m free space reference distance that is physically based, thus allowing easy use for varying distances without a calculator, through the use of just a single parameter (PLE, or $n$). The ABG model has three parameters that vary wildly across different scenarios and frequency ranges, and the AB model parameters vary widely across frequencies and distances, while reducing the standard deviation by only a fraction of a dB compared to the simpler, physically-based CI model. The improvement in error with the more complex three-parameter ABG model is insignificant --- usually well below an order of magnitude of the actual standard deviation value of all models considered here.

The results suggest that the CI and ABG models offer very comparable modeling performance using real data, with the CI model offering simplicity and a physical basis with one parameter, and providing a more conservative NLOS path loss estimate at large distances, while the ABG model offers a fraction of a dB smaller SF and requires three parameters that are not physically based, while predicting less path loss close to the transmitter and more loss at greater distances. While the CI model offers virtually identical modeling accuracy and simplicity using only one model parameter (the PLE), the ABG model offers slightly improved accuracy at the expense of three parameters. Furthermore, the CI model has a very similar form compared to the existing 3GPP path loss model, as one merely needs to replace the floating constant, which has been shown to vary widely across different measurements, frequencies and scenarios, by the free-space path loss that is a function of frequency based on a 1 m standard close-in reference distance. This subtle change leads to much easier analysis, stability, and accuracy over a vast range of microwave and mmWave frequencies, distances, and scenarios, while using a simpler model with fewer parameters.


\section*{Appendix}
Mathematical derivations for the closed-form solutions for the ABG and CI models, by solving for model parameters that minimize the SF standard deviation, are provided in this appendix.
\subsection{ABG Path Loss Model}
The ABG model can be expressed as (with 1 m reference distance and 1 GHz reference frequency)\cite{Pie12}:
\begin{equation}\label{ABG}
\begin{split}
\PL^{\ABG}(f,d)[\dB]=&10\alpha \log_{10}(\frac{d}{1~m})+\beta+10\gamma \log_{10}(\frac{f}{1~GHz})\\
&+\chi_{\sigma}^{\ABG}
\end{split}
\end{equation}

\noindent Assuming $B=\PL^{ABG}(f,d)[\dB]$, $D=10\log_{10}(d)$, and $F=10\log_{10}(f)$ in~\eqref{ABG}, the SF is given by:
\begin{equation}
\chi_{\sigma}^{\ABG}=B-\alpha D-\beta-\gamma F
\end{equation}

\noindent Then the SF standard deviation is:
\begin{equation}\label{ABG_sigma}
\sigma^{\ABG} = \sqrt{\sum{{\chi_{\sigma}^{\ABG}}^2}/N}=\sqrt{\sum{(B-\alpha D-\beta-\gamma F)^2}/N}
\end{equation}

Minimizing the fitting error is equivalent to minimizing $\sum{(B-\alpha D-\beta-\gamma F)^2}$, which means its partial derivatives with respect to $\alpha$, $\beta$, and $\gamma$ should be zero, as shown by~\eqref{ABG_derA},~\eqref{ABG_derB}, and~\eqref{ABG_derG}.
\begin{equation}\label{ABG_derA}
\begin{split}
\frac{\partial \sum{(B-\alpha D-\beta-\gamma F)^2}}{\partial \alpha}=&2(\alpha\sum{D^2}+\beta\sum{D}\\
&+\gamma\sum{DF}-\sum{DB})\\
=&0
\end{split}
\end{equation}

\begin{equation}\label{ABG_derB}
\begin{split}
\frac{\partial \sum{(B-\alpha D-\beta-\gamma F)^2}}{\partial \beta}=&2(\alpha\sum{D}+N\beta+\gamma\sum{F}\\
&-\sum{B})\\
=&0
\end{split}
\end{equation}

 \begin{equation}\label{ABG_derG}
\begin{split}
\frac{\partial \sum{(B-\alpha D-\beta-\gamma F)^2}}{\partial \gamma}=&2(\alpha\sum{DF}+\beta\sum{F}\\
&+\gamma\sum{F^2}-\sum{FB})\\
=&0
\end{split}
\end{equation}

\begin{figure*}
\begin{equation}\label{A}
\alpha=\frac{(\sum{D}\sum{B}-N\sum{DB})((\sum{F})^2-N\sum{F^2})-(\sum{D}\sum{F}-N\sum{DF})(\sum{F}\sum{B}-N\sum{FB})}{((\sum{D})^2-N\sum{D^2})((\sum{F})^2-N\sum{F^2})-(\sum{D}\sum{F}-N\sum{DF})^2}
\end{equation}
\end{figure*}

\begin{figure*}
\begin{equation}\label{B}
\beta=\frac{(\sum{D}\sum{FB}-\sum{B}\sum{DF})(\sum{F}\sum{D^2}-\sum{D}\sum{DF})-(\sum{B}\sum{D^2}-\sum{D}\sum{DB})(\sum{D}\sum{F^2}-\sum{F}\sum{DF})}{((\sum{D})^2-N\sum{D^2})(\sum{D}\sum{F^2}-\sum{F}\sum{DF})+(\sum{D}\sum{F}-N\sum{DF})(\sum{F}\sum{D^2}-\sum{D}\sum{DF})}
\end{equation}
\end{figure*}

\begin{figure*}
\begin{equation}\label{G}
\gamma=\frac{(\sum{F}\sum{B}-N\sum{FB})((\sum{D})^2-N\sum{D^2})-(\sum{D}\sum{F}-N\sum{DF})(\sum{D}\sum{B}-N\sum{DB})}{((\sum{F})^2-N\sum{F^2})((\sum{D})^2-N\sum{D^2})-(\sum{D}\sum{F}-N\sum{DF})^2}
\end{equation}
\end{figure*}

\noindent It is found from~\eqref{ABG_derA},~\eqref{ABG_derB}, and~\eqref{ABG_derG} that
\begin{equation}\label{eqA}
\alpha\sum{D^2}+\beta\sum{D}+\gamma\sum{DF}-\sum{DB}=0
\end{equation}

\begin{equation}\label{eqB}
\alpha\sum{D}+N\beta+\gamma\sum{F}-\sum{B}=0
\end{equation}

\begin{equation}\label{eqG}
\alpha\sum{DF}+\beta\sum{F}+\gamma\sum{F^2}-\sum{FB}=0
\end{equation}

\noindent Through calculation and simplification, we obtain the closed-form solutions for $\alpha$, $\beta$, and $\gamma$ as shown by~\eqref{A},~\eqref{B}, and~\eqref{G}, respectively. Finally, the minimum SF standard deviation for the ABG model can be obtained by plugging~\eqref{A},~\eqref{B}, and~\eqref{G} back into~\eqref{ABG_sigma}.

\subsection{CI Path Loss Model}\label{subsec:CI}
The expression for the CI model with a reference distance of 1 m is given by\cite{Rap15:TCOM}:
\begin{equation}\label{CI}
\PL^{\CI}(f,d)[\dB]=\FSPL(f, 1~m)[\dB]+10n\log_{10}(d)+\chi_{\sigma}^{\CI}
\end{equation}

\noindent Thus the SF is:
\begin{equation}
\begin{split}
\chi_{\sigma}^{\CI}&=\PL^{\CI}(f,d)[\dB]-\FSPL(f, 1~m)[\dB]-10n\log_{10}(d)\\
&=A-nD
\end{split}
\end{equation}

\noindent where $A$ represents $\PL^{CI}(f,d)[\dB]-\FSPL(f, 1~m)[\dB]$, and $D$ denotes $10\log_{10}(d)$. Then the SF standard deviation is:
\begin{equation}
\sigma^{\CI} = \sqrt{\sum{{\chi_{\sigma}^{\CI}}^2}/N}=\sqrt{\sum{(A-nD)^2}/N}
\end{equation}

\noindent where $N$ is the number of path loss data points. Thus minimizing the SF standard deviation $\sigma^{\CI}$ is equivalent to minimizing the term $\sum{(A-nD)^2}$. When $\sum{(A-nD)^2}$ is minimized, its derivative with respect to $n$ should be zero, i.e.,
\begin{equation}\label{PLE_der}
\begin{split}
\frac{d\sum{(A-nD)^2}}{dn}&=\sum{2D(nD-A)}=0\\
\end{split}
\end{equation}

\noindent Therefore, from~\eqref{PLE_der} we have
\begin{equation}\label{PLE}
n=\frac{\sum{DA}}{\sum{D^2}}
\end{equation}



\ifCLASSOPTIONcaptionsoff
  \newpage
\fi

\bibliographystyle{IEEEtran}
\bibliography{bibliography}

\begin{thebibliography}{10}
\providecommand{\url}[1]{#1}
\csname url@samestyle\endcsname
\providecommand{\newblock}{\relax}
\providecommand{\bibinfo}[2]{#2}
\providecommand{\BIBentrySTDinterwordspacing}{\spaceskip=0pt\relax}
\providecommand{\BIBentryALTinterwordstretchfactor}{4}
\providecommand{\BIBentryALTinterwordspacing}{\spaceskip=\fontdimen2\font plus
\BIBentryALTinterwordstretchfactor\fontdimen3\font minus
  \fontdimen4\font\relax}
\providecommand{\BIBforeignlanguage}[2]{{%
\expandafter\ifx\csname l@#1\endcsname\relax
\typeout{** WARNING: IEEEtran.bst: No hyphenation pattern has been}%
\typeout{** loaded for the language `#1'. Using the pattern for}%
\typeout{** the default language instead.}%
\else
\language=\csname l@#1\endcsname
\fi
#2}}
\providecommand{\BIBdecl}{\relax}
\BIBdecl

\bibitem{Rap13:Access}
\text{T. S. Rappaport} \textit{et al.}, ``Millimeter wave mobile communications
  for \text{5G} cellular: It will work!'' \emph{IEEE Access}, vol.~1, pp.
  335--349, 2013.

\bibitem{And14}
\text{J. G. Andrews} \textit{et al.}, ``What will 5\text{G} be?'' \emph{IEEE
  Journal on Selected Areas in Communications}, vol.~32, no.~6, pp. 1065--1082,
  Jun. 2014.

\bibitem{Boc14}
\text{F. Boccardi} \textit{et al.}, ``Five disruptive technology directions for
  5\text{G},'' \emph{IEEE Communications Magazine}, vol.~52, no.~2, pp. 74--80,
  Feb. 2014.

\bibitem{Vio88}
\text{E. J. Violette} \textit{et al.}, ``Millimeter-wave propagation at street
  level in an urban environment,'' \emph{IEEE Transactions onGeoscience and
  Remote Sensing}, vol.~26, no.~3, pp. 368--380, 1988.

\bibitem{Lov94}
\text{G. Lovnes} \textit{et al.}, ``Channel sounding measurements at 59 ghz in
  city streets,'' in \emph{5th IEEE International Symposium on Personal, Indoor
  and Mobile Radio Communications}, Sep. 1994, pp. 496--500 vol.2.

\bibitem{Smu97}
P.~Smulders and L.~Correia, ``Characterisation of propagation in 60 ghz radio
  channels,'' \emph{Electronics Communication Engineering Journal}, vol.~9,
  no.~2, pp. 73--80, Apr. 1997.

\bibitem{Hur14}
\text{S. Hur} \textit{et al.}, ``Synchronous channel sounder using horn antenna
  and indoor measurements on 28 ghz,'' in \emph{2014 IEEE International Black
  Sea Conference on Communications and Networking (BlackSeaCom)}, May 2014, pp.
  83--87.

\bibitem{Roh14}
\text{W. Roh} \textit{et al.}, ``Millimeter-wave beamforming as an enabling
  technology for 5g cellular communications: theoretical feasibility and
  prototype results,'' \emph{IEEE Communications Magazine}, vol.~52, no.~2, pp.
  106--113, Feb. 2014.

\bibitem{Kyro12}
M.~Kyro, V.~Kolmonen, and P.~Vainikainen, ``Experimental propagation channel
  characterization of mm-wave radio links in urban scenarios,'' \emph{IEEE
  Antennas and Wireless Propagation Letters}, vol.~11, pp. 865--868, 2012.

\bibitem{Rap15:TCOM}
T.~S. Rappaport \emph{et~al.}, ``Wideband millimeter-wave propagation
  measurements and channel models for future wireless communication system
  design ({I}nvited {P}aper),'' \emph{IEEE Transactions on Communications},
  vol.~63, no.~9, pp. 3029--3056, Sep. 2015.

\bibitem{Rap13:TAP}
\text{T. S. Rappaport} \textit{et al.}, ``Broadband millimeter-wave propagation
  measurements and models using adaptive-beam antennas for outdoor urban
  cellular communications,'' \emph{IEEE Transactions on Antennas and
  Propagation}, vol.~61, no.~4, pp. 1850--1859, April 2013.

\bibitem{Mac14:PIMRC}
\text{G. R. MacCartney, Jr.} \textit{et al.}, ``Omnidirectional path loss
  models from measurements recorded in new york city at \text{28 GHz and 73
  GHz},'' in \emph{IEEE International Symposium on Personal Indoor and Mobile
  Radio Communications (PIMRC)}, Sep. 2014.

\bibitem{GRM13:Globecom}
------, ``Path loss models for 5{G} millimeter wave propagation channels in
  urban microcells,'' in \emph{IEEE Global Communications Conference
  (GLOBECOM)}, Dec. 2013, pp. 3948--3953.

\bibitem{Pie12}
\text{S. Piersanti} \textit{et al.}, ``Millimeter waves channel measurements
  and path loss models,'' in \emph{2012 IEEE International Conference on
  Communications (ICC)}, Jun. 2012, pp. 4552--4556.

\bibitem{And95}
J.~B. Andersen, T.~S. Rappaport, and S.~Yoshida, ``Propagation measurements and
  models for wireless communications channels,'' \emph{IEEE Communications
  Magazine}, vol.~33, no.~1, pp. 42--49, Jan. 1995.

\bibitem{WINNER}
\text{P. Kyosti, \textit{et al.}}, ``\text{WINNER II channel models},''
  \emph{European Commission, \text{IST-4-027756-WINNER, Tech. Rep. D1.1.2}},
  Sep. 2007.

\bibitem{Rappaport:Wireless2nd}
T.~S. Rappaport, \emph{Wireless Communications: Principles and Practice},
  2nd~ed.\hskip 1em plus 0.5em minus 0.4em\relax Upper Saddle River, NJ:
  Prentice Hall, 2002.

\bibitem{Tho16:VTC}
\text{Timothy A. Thomas,~\textit{et al.}}, ``A prediction study of path loss
  models from 2-73.5 {GHz} in an urban-macro environment,''
  \emph{\textit{submitted to 2016 IEEE 83rd Vehicular Technology Conference
  (VTC Spring)}}, May 2016.

\bibitem{Deng15}
S.~Deng, M.~K. Samimi, and T.~S. Rappaport, ``28 \text{GHz} and 73 \text{GHz}
  millimeter-wave indoor propagation measurements and path loss models,'' in
  \emph{2015 IEEE International Conference on Communications (ICC), ICC
  Workshops}, Jun. 2015.

\bibitem{Hata:TVT80}
M.~Hata, ``Empirical formula for propagation loss in land mobile radio
  services,'' \emph{IEEE Transactions on Vehicular Technology}, vol.~29, no.~3,
  pp. 317--325, Aug. 1980.

\bibitem{Akd14}
M.~R. Akdeniz \emph{et~al.}, ``Millimeter wave channel modeling and cellular
  capacity evaluation,'' \emph{IEEE Journal on Selected Areas in
  Communications}, vol.~32, no.~6, pp. 1164--1179, June 2014.

\bibitem{3GPP:25996}
\text{3GPP TR 25.996}, ``Spatial channel model for multiple input multiple
  output \text{(MIMO)} simulations,'' Sep. 2012.

\bibitem{3GPP:36814}
\text{3GPP TR 36.814}, ``Further advancements for \text{E-UTRA} physical layer
  aspect,'' Mar. 2010.

\end{thebibliography}

\end{document}